\documentclass[10pt, twocolumn, pre, aps, superscriptaddress, showpacs]{revtex4-1}
\usepackage{amsmath, graphicx, subfigure, tikz}
\begin{document}

\title{Asymmetric Phase Diagrams, Algebraically Ordered BKT Phase, and \\
 Peninsular Potts Flow Structure in Long-Range Spin Glasses}
    \author{S. Efe G\"{u}rleyen}
    \affiliation{Department of Physics, Istanbul Technical University, Maslak, Istanbul 34469, Turkey}
    \author{A. Nihat Berker}
    \affiliation{Faculty of Engineering and Natural Sciences, Kadir Has University, Cibali, Istanbul 34083, Turkey}
    \affiliation{Department of Physics, Massachusetts Institute of Technology, Cambridge, Massachusetts 02139, USA}

\begin{abstract}

The Ising spin-glass model on the three-dimensional $(d=3)$ hierarchical lattice with long-range ferromagnetic or spin-glass interactions is studied by the exact renormalization-group solution of the hierarchical lattice.  The chaotic characteristics of the spin-glass phases are extracted in the form of our calculated, in this case continuously varying, Lyapunov exponents.  Ferromagnetic long-range interactions break the usual symmetry of the spin-glass phase diagram.  This phase-diagram symmetry-breaking is dramatic, as it is underpinned by renormalization-group peninsular flows of the Potts multicritical type.  A Berezinski-Kosterlitz-Thouless (BKT) phase with algebraic order and a BKT-spinglass phase transition with continuously varying critical exponents are seen.  Similarly, for spin-glass long-range interactions, the Potts mechanism is also seen, by the mutual annihilation of stable and unstable fixed distributions causing the abrupt change of the phase diagram.  On one side of this abrupt change, two distinct spin-glass phases, with finite (chaotic) and infinite (chaotic) coupling asymptotic behaviors are seen with a spin-glass-to-spin-glass phase transition.
\end{abstract}
\maketitle
\section{Introduction: Long-Range Spin-Glasses}
Spin-glass systems \cite{EdwardsAnderson}, composed of frozen randomly distributed competing interactions, such as ferromagnetic and antiferromagnetic interactions or, more recently \cite{Caglar1, Caglar2, Caglar3}, left- and right-chiral (i.e., helical \cite{Ostlund,Surface3}) interactions, exhibit phases with distinctive spin-glass order.  A prime characteristic of the spin-glass phase is the chaotic behavior \cite{McKayChaos,McKayChaos2,BerkerMcKay,Hartford,ZZhu,Katzgraber3,Fernandez,Fernandez2,Eldan,Wang2,Parisi3} of the effective temperature under scale change, which also means the major changes of the macroscopic properties under minor changes of the external paramater such as temperature.\cite{Aral}  In this study, we consider the spin-glass system of Ising spins on a three-dimensional $(d=3)$ hierarchical lattice \cite{BerkerOstlund,Kaufman1,Kaufman2}, with the inclusion of long-range interactions \cite{Hinczewski,percolation,Jiang}.  We study, in turn, ferromagnetic and spin-glass long-range interactions.  Much qualitatively new behavior emerges from the inclusion of these long-range interactions.  Refs. \cite{Derevyagin2,Chio,Teplyaev,Myshlyavtsev,Derevyagin,Shrock,Monthus,Sariyer} are recent works using exactly soluble hierarchical models.
\begin{figure}[ht!]
\centering
\includegraphics[scale=0.29]{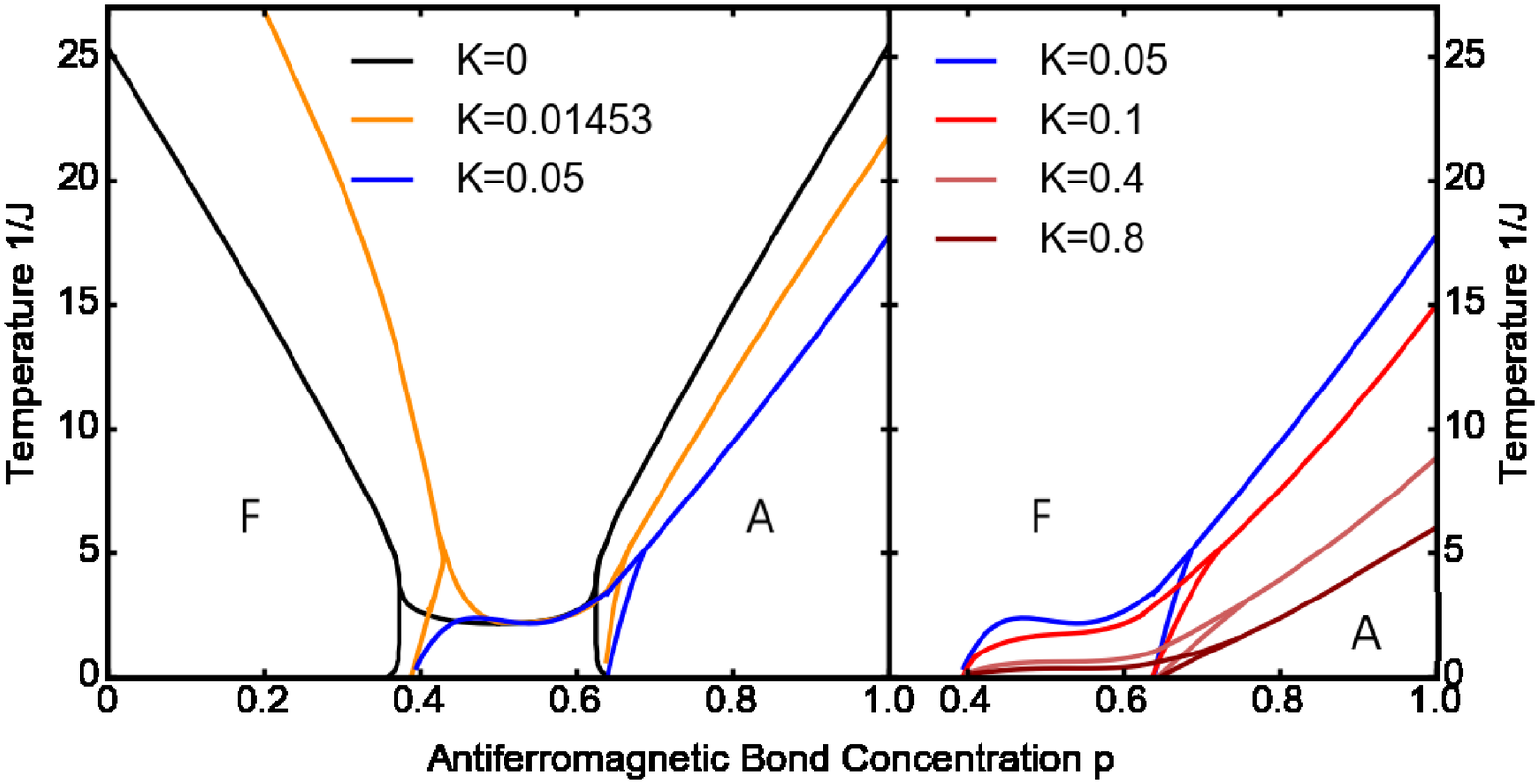}
\caption{Calculated phase diagrams of the Ising spin glass with long-range ferromagnetic interaction $K$ in $d=3$. In the left panel, the phase diagram that starts leftmost is for $K=0$, no long-range interaction, and is the standart spin-glass phase diagram with ferromagnetic-antiferromagnetic symmetry about the $p=0.5$ line.  The ferromagnetic and antiferromagnetic phases are marked respectively as F and A.  Between these phases, there are the spin-glass and disordered phases, respectively at low and high temperature.  In the next phase diagram to the right in the left panel, for long-range ferromagnetic interaction $K=0.01453$, the phase diagram is slightly deformed and loses the ferromagnetic-antiferromagnetic symmetry.  For $K>0$, the disordered phase is replaced by a Berezinski-Kosterlitz-Thouless (BKT) phase with algebraic order.  At $K=0.01453$, the BKT phase precipitously disappears, by the renormalization-group mechanism of the peninsular Potts flows, explained in the text and in Fig. 3. For $K>0.01453$, there is a direct phase transition between the ferromagnetic and antiferromagnetic phases, as seen for $K=0.05$, the rightmost phase diagram in the left panel of this figure. In the right panel of this figure, the evolution of this phase diagram is seen from the phase diagrams for $K=0.05,0.1,0.4,0.8$, from top to bottom.}
\end{figure}

Our model, with nearest-neighbor spin-glass interactions and long-range ferromagnetic or spin-glass interactions, is defined by the Hamiltonian
\begin{equation}
-\beta \mathcal{H}=\sum_{\langle ij \rangle} J_{ij} s_i s_j \,+\sum_{LR} K_{ij} s_i s_j \,,
\end{equation}
where $\beta=1/kT$, $s_i = \pm1$ at each site $i$ of the lattice, and the sum $\langle ij \rangle$ is over all pairs of
nearest-neighbor sites.  The bond $J_{ij}$ is ferromagnetic $+J>0$ or antiferromagnetic $-J$ with probabilities $1-p$ and $p$, respectively. The long-range interaction $LR$ is between all spins pairs beyond the first neighbors.  We have studied the two cases where, for all further-neighbor spin pairs, the long-range interaction is (a) ferromagnetic $K_{ij}=K>0$ or (b) frozen ferromagnetic or antiferromagnetic $K_{ij}=\pm K$ with equal probability, namely a spin-glass interaction.  By symmetry, and a simple reflection (which is meaningful, as the phase diagrams become asymmetric) of the phase diagrams about the $p=0.5$ line, case (a) is equivalent to antiferromagnetic $K_{ij}=-K<0$ interaction for all further-neighbor spin pairs.  As seen in Fig.1, the introduction of long-range interaction qualitatively affects the system, introducing a new phase (the BKT phase) and a new mechanism of phase collapse (the peninsular Potts flow renormalization-group mechanism).
\begin{figure}[ht!]
\centering
\includegraphics[scale=0.22]{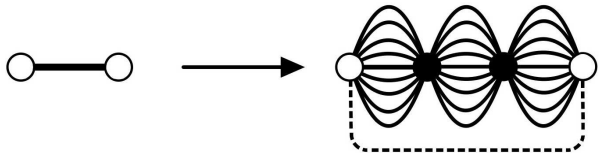}
\caption{The hierachical lattice is constructed by the repeated self-imbedding of a graph into bond. \cite{BerkerOstlund}  The dashed line in the graph represents the long-range interaction.  The renormalization-group exact solution proceeds in the opposite direction:  Summation over the spins on the internal sites (full circles) of the graph gives the renormalized interaction $J'$ between the spins on the external sites (open circles) as a function of the interactions $J$ and $K$ on the graph, namely the recursion relation $J'=J'(\{J_{ij}\},K)$, where $\{J_{ij}\}$ are the nearest-neighbor interactions, in general with different values, inside the graph.}
\end{figure}
\section{Construction of the Hierarchical Lattice and Renormalization-Group Exact Solution}
The construction of the hierarchical lattice is explained in Fig. 2.  The number (in this case 27) of nearest-neighbor interactions replacing a single nearest-neighbor interaction gives the dimensionality as $b^d$, where $b$ is the length-rescaling factor, namely the number of bonds in the shortest path between the external sites of the graph.  In the present case, $b=3$ and therefore $d=3$.

The renormalization-group transformation is effected by expressing the nearest-neighbor interaction as a $2 \times 2$ transfer matrix, $T_{ij}(s_i,s_j)= e^{E_{ij}(s_i,s_j)}$, where the energy $E_{ij}(s_i,s_j)$ is initially as given in the first term of Eq.(1). For each renormalization-group trajectory, initially 4000 unrenormalized transfer matrices $\{T_{ij}\}$ are generated randomly from the double-delta distribution characterized by the probability $p$ as explained above.  In each consecutive renormalization-group transformation, a new (renormalized) set of 4000 transfer matrices $\{T'_{ij}\}$ is generated, using the recursion relation explained in Fig. 2 and in(A-G) below,  randomly choosing each of the $b^d$ unrenormalized transfer matrices $T_{ij}$ inside the graph from the 4000 transfer matrices generated from the previous renormalization-group transformation.  Thus, a renormalization-group flow of the quenched probability distribution of the interactions \cite{AndelmanBerker} is obtained.

The generation of a set of renormalized transfer matrices is broken into binary steps \cite{Ilker1,Ilker2,Ilker3} that accomplish the dictate of Fig. 2:

(A) First, the starting set of tranfer matrices is combined with itself, by randomly chosing two transfer matrices, $\mathbf{T^{(1)}}$ and $\mathbf{T^{(2)}}$, from the set and multiplying matrix elements at each position, $T^{(1)}_{ij}*T^{(2)}_{ij}$, thus obtaining a new transfer matrix.  4000 such new matrices are generated.

(B) The set generated in (A) is combined with itself, using the procedure described in (A).

(C) The set generated in (B) is combined with itself, using the procedure described in (A).

(D) The set generated in (C) is combined with the initial set used in (A), using the procedure described in (A). This completes the combination of $b^{d-1}=9$ parallel bonds shown in each bubble in Fig. 2.

(E) The set generated in (D) is combined with itself, by randomly chosing two transfer matrices, $\mathbf{T^{(1)}}$ and $\mathbf{T^{(2)}}$, from the set and matrix multiplying, $\mathbf{T^{(1)} \cdot T^{(2)}}$.

(F) The set generated in (E) is combined with the initial set used in (E), using the procedure described in (E). This completes the elimination of the internal sites in Fig. 2 by decimation.

(G) The anti-diagonals of each transfer matrix in the set are multiplied by $\exp {-2K}$.

This also completes the renormalization-group transformation, obtaining the set of 4000 renormalized transfer matrices $\{\mathbf{T'}\}$ from the set of 4000 unrenormalized transfer matrices $\{\mathbf{T}\}$.  This renormalization group-transformation is repeated many times to obtain a renormalization-group trajectory of the quenched probability distribution.

With no loss of generality, each time that a transfer matrix is constructed as described in the previous paragraphs, the matrix elements are divided by the largest element, so that eventually all matrix elements are between 1 and 0, inclusive.  This allows the repetition of the renormalization-group transformation as much as necessary (in practice, thousands of times) without running into numerical overflow problems, needed for the determination of thermodynamic phase sinks, runaway exponents, and the Lyapunov exponents of chaos.

For trajectories starting at $(J,K,p)$ in the ferromagnetic phase, all transfer matrices in the set asymptotically renormalize to 1 in the diagonals and 0 in the anti-diagonals.  For trajectories starting at $(J,K,p)$ in the antiferromagnetic phase, all transfer matrices in the set asymptotically renormalize to 0 in the diagonals and 1 in the anti-diagonals.  For trajectories starting at $(J,K,p)$ in the spin-glass phase, all transfer matrices in the set asymptotically renormalize to 1 in the diagonals or anti-diagonal randomly, simultaneously with 0 in the anti-diagonals or diagonals.  For trajectories starting in the algebraically ordered BKT phase, all transfer matrices in the set asymptotically renormalize to 1 in the diagonals and to a value between 1 and 0 in the anti-diagonals, continuously varying based on the inital $(J,K,p)$ of the trajectory.  For the trajectories starting in the disordered phase, all transfer matrices in the set renormalize to 1 in the diagonals and anti-diagonals.  Phase boundaries in $(J,K,p)$ are obtained by numerically determining the boundaries of these different asymptotic behaviors.
\begin{figure}[ht!]
\centering
\includegraphics[scale=0.23]{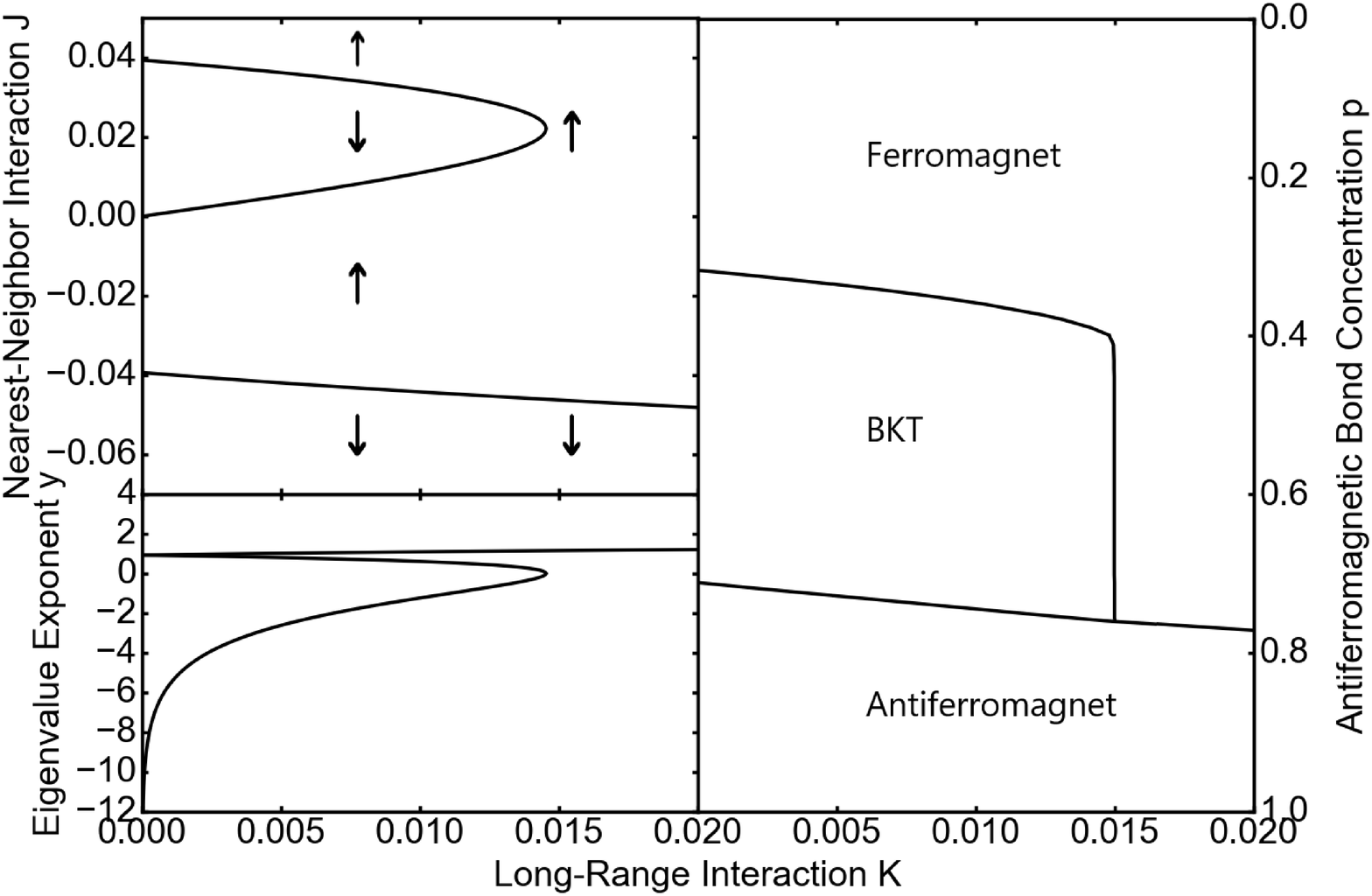}
\caption{The peninsular Potts renormalization-group flow mechanism and the precipitous phase diagram.  In the lower left panel, the upper line gives the eigenvalue exponents $y$ for the phase transitions from the antiferromagnetic phase.  The positive part of the lower curve gives the eigenvalue exponents for the phase transitions between the algebraically ordered and ferromagnetic phases.  The phase diagram on the right is calculated at constant temperature $J^{-1} = 0.1$. Because of the Potts-peninsular mechanism, explained in the Sec. III, part of the phase boundary between the ferromagnetic and BKT phases should be and is calculated to be vertical here.}
\end{figure}
\section{Potts-Peninsular Renormalization-Group Mechanism and Precipitous Phase Diagram}
Quenched randomness amplifies in renormalization-group trajectories starting in the spin-glass phase and shows chaotic rescaling behavior.  Quenched randomness deamplifies in renormalization-group trajectories starting in the four other phases.  In this case, the recursion relation constructed in the previous section becomes
\begin{equation}
J' = \tanh^{-1}\{[\tanh(9J)]^3\} + K \,.
\end{equation}
Solving Eq.(2) for $J'=J \equiv J^*$ gives the fixed point interactions $J^*$ as a function of $K$, shown in the upper right panel of Fig. 3.  Taking the derivative of Eq.(2) at the fixed point,
\begin{equation}
\frac {dJ'}{dJ} = \frac {27[\tanh(9J)]^2}{1+[\tanh(9J)]^2+[\tanh(9J)]^4} = b^y \,,
\end{equation}
the eigenvalue exponents $y$ at the fixed point are obtained.  These are shown in the lower left panel of Fig. 3.

The peninsular Potts renormalization-group flow mechanism and the precipitous phase diagram are given in Fig. 3.  The upper left panel shows the lines of fixed points as a function of the long-range interaction $K$, calculated from Eq.(2).  This calculation is done in the non-random limit where all renormalization-group trajectories flow, from phases outside the spin-glass phase.  In this upper left panel, the lower curve is the fixed line, unstable to the renormalization-group flows, giving the phase boundary between the antiferromagnetic phase and, for $K<0.01453$ where the upper flows hit the stable branch of the peninsula, the BKT phase and, for $K>0.01453$ where the upper flows miss the peninsula beyond its tip, the ferromagnetic phase.  Therefore, the BKT phase precipitously disappears for at $K=0.01453$.  Due to this catastrophic changeover \cite{Thom}, in Fig. 3, part of the phase boundary between the ferromagnetic and BKT phases should be and is calculated to be vertical.

In the lower left panel of Fig. 3, the lower branch of the peninsula is a fixed line, stable to the renormalization-group flows, constituting the sink of the algebraically ordered BKT phase.  The upper branch of the peninsula is a fixed line, unstable to the flows, giving the phase transition between the BKT phase and the ferromagnetic phase.  The renormalization-group flows are indicated with the arrows.  The flows at the upper and lower edges of the panel proceed to $J=+\infty$ and $J=-\infty$, constituting the sinks of the ferromagnetic and antiferromagnetic phases respectively.  The unstable fixed lines give the phase transitions.  As seen in the lower left panel of Fig. 3, the eigenvalue exponent $y$ and therefore the critical exponents (e.g., the correlation-length critical exponent $\nu$) vary continuously along the phase boundaries. This peninsular renormalization-group flow mechanism previously has only been seen in Potts models in $d$ dimensions, realizing the changeover from second- to first-order phase transitions of the Potts models.\cite{spinS7,Nienhuis1,Nienhuis2,AndelmanPotts0,AndelmanPotts1,Nienhuis3}
\begin{figure}[ht!]
\centering
\includegraphics[scale=0.2]{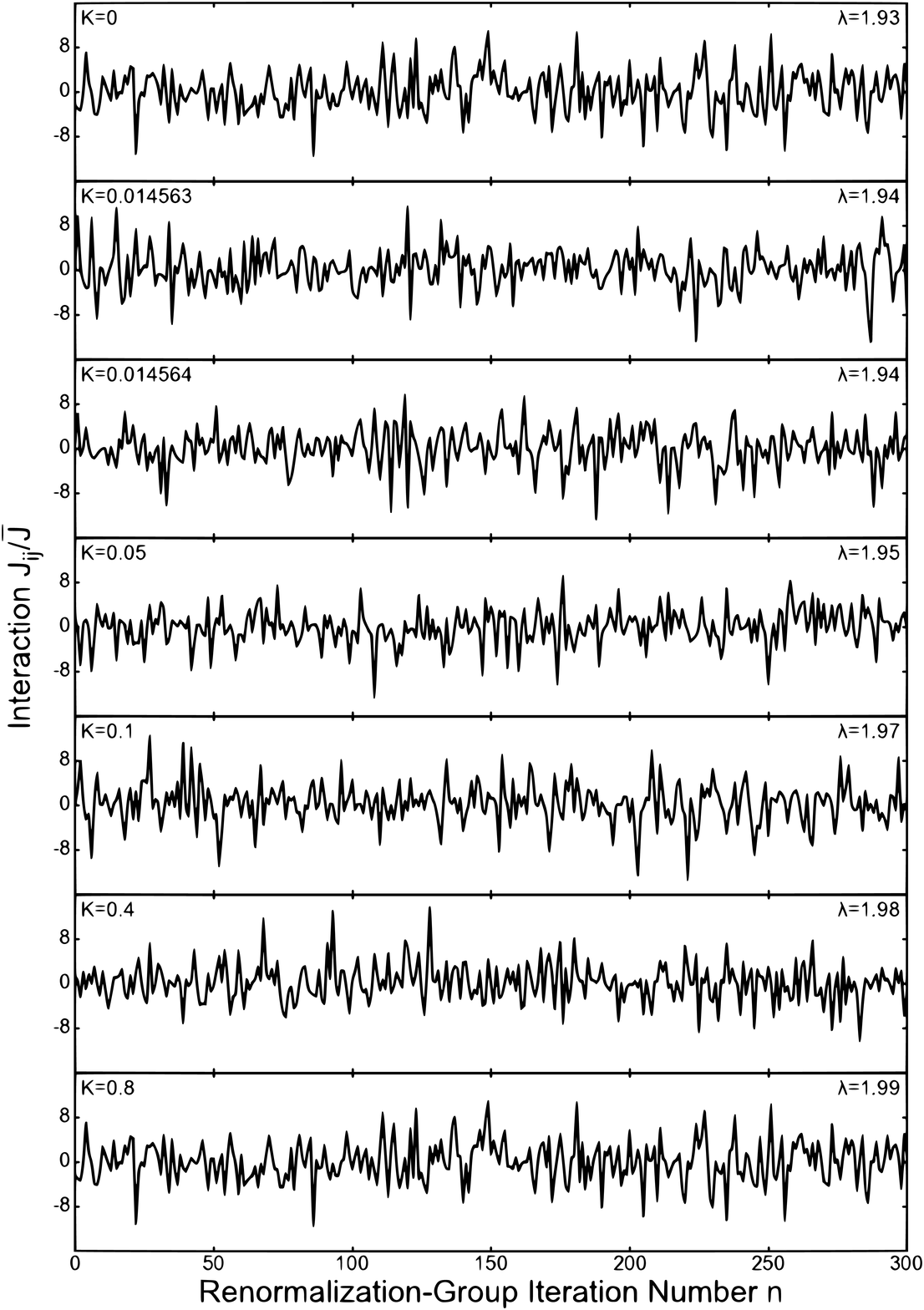}
\caption{The chaotic renormalization-group trajectory of the interaction $J_{ij}$ at a given location $<ij>$, for various long-range interactions $K$. The calculated Lyapunov exponents $\lambda$ are also given and increase with ferromagnetic long-range interaction $K$.  The calculated runaway exponent is $y_R=0.24$, showing simultaneous strong-chaos and strong-coupling behaviors.}
\end{figure}
\section{Asymmetric Phase Diagrams with Algebraically Ordered Berezinski-Kosterlitz-Thouless Phase}
Calculated phase diagrams of the Ising spin glass with long-range ferromagnetic interaction $K$ in $d=3$ are shown in Fig. 1. In the left panel, the phase diagram that starts leftmost is for $K=0$, no long-range interaction, and is the standart spin-glass phase diagram with ferromagnetic-antiferromagnetic symmetry about the $p=0.5$ line.  The ferromagnetic and antiferromagnetic phases are marked respectively as F and A.  Between these phases, there are the spin-glass and disordered phases, respectively at low and high temperature.  In the next phase diagram to the right, for long-range ferromagnetic interaction $K=0.01453$, the phase diagram is slightly deformed and loses the ferromagnetic-antiferromagnetic symmetry.  For $K>0$, the disordered phase is replaced by a Berezinski-Kosterlitz-Thouless (BKT) phase with algebraic order. This phase has algebraic order, since its sink line continuously varies and is at non-zero and non-infinite interactions.  In general, the correlation length at a fixed point is either zero, or infinite, due to the scale-free nature of this point.  In the present case, the zero option is eliminated by the fixed-point interactions being non-zero and non-infinite.  Therefore, the BKT attractive fixed line (phase sink) and all points flowing to it under renormalization group have infinite correlation length and algebraic order.\cite{Kosterlitz,Jose,BerkerNelson,BerkerKadanoff1,BerkerKadanoff2}

At $K=0.01453$, the BKT phase precipitously disappears, by the renormalization-group mechanism of the peninsular Potts flows, explained in Sec. III and in Fig. 3.  Thus, our phase diagram calculations (Fig. 1) with global renormalization-group flows exactly yield and confirm the peninsular tip obtained from the fixed-point calculation using Eq. (2) (Fig. 3).  For $K>0.01453$, there is a direct phase transition between the ferromagnetic and antiferromagnetic phases, as seen for $K=0.05$, the rightmost phase diagram in the left panel of Fig. 1. In the right panel of Fig. 1, the evolution of this phase diagram is seen from the phase diagrams for $K=0.05,0.1,0.4,0.8$, from top to bottom.
\section{Chaos Continuously Varying within the Spin-Glass Phase: Lyapunov Exponent and Runaway Exponent}
The spin-glass phase is a phase induced by competing quenched randomness and that does not otherwise exist.  The competing interactions can be ferromagnetic versus antiferromagnetic, as here, or left- and right-chiral interactions.  A distinctive characteristic of the spin-glass phase is chaos under scale change \cite{McKayChaos}.  In the present work, the asymptotic chaotic trajectory continuously varies quantitatively with the long-range interaction $K$.

The asymptotically chaotic renormalization-group trajectories starting within the spin-glass phase are shown for various values of the long-range interaction $K$ in Fig. 4, where, for each $K$, the consecutively renormalized (combining with neighboring interactions) values at a given location $<ij>$ are followed. The strength of chaos is measured by the Lyapunov exponent \cite{Collet,Hilborn}
\begin{equation}
\lambda = \lim _{n\rightarrow\infty} \frac{1}{n} \sum_{k=0}^{n-1}
\ln \Big|\frac {dx_{k+1}}{dx_k}\Big|\,,
\end{equation}
where $x_k = J_{ij}/\overline{J}$ at step $k$ of the renormalization-group trajectory and $\overline{J}$ is the average of the absolute value of the interactions in the quenched random distribution.  We calculate the Lyapunov exponents by discarding the first 1000 renormalization-group steps (to eliminate crossover from initial conditions to asymptotic behavior) and then using the next 9000 steps.  For a given $K$ value, the initial $(J,p)$ values do not matter, as long as they are within the spin-glass phase.  In the absence of long-range interaction, $K=0$, the Lyapunov exponent is calculated to be $\lambda = 1.93$, as in previous work \cite{Ilker2,ArtunBerker}.  With increasing long-range ferromagnetic interaction, the Lyapunov exponent and therefore chaos increase, to the value of $\lambda = 1.99$ for $K=0.8$.

In addition to chaos, the renormalization-group trajectories show asymptotic strong coupling behavior,
\begin{equation}
\overline{J'} = b^{y_R}\, \overline{J}\,,
\end{equation}
where $y_R >0$ is the runaway exponent \cite{Demirtas}.  Again using 9000 renormalization-group steps after discarding 1000 steps, we find $y_R =0.24$ for all values of $K$. In fact, $y_R =0.24$ was also found previously for all values of the spin $s$ \cite{ArtunBerker}.
\begin{figure}[ht!]
\centering
\includegraphics[scale=0.5]{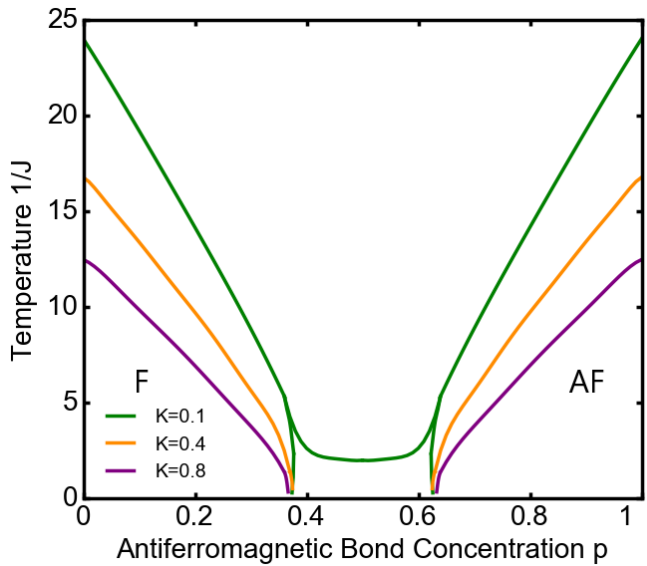}
\caption{Calculated phase diagrams of the Ising spin glass with long-range spin-glass interaction $\pm K$ in $d=3$. From top to bottom, the phase diagrams are for $K=0.1,0.4,0.8$. The ferromagnetic and antiferromagnetic phases are marked respectively as F and A.  Between these phases, for $K=0.1$, there are the weak-coupling and strong-coupling spin-glass phases, respectively at high and low temperature.  The weak-coupling spin-glass phase occurs for $0 < K < 0.1883$ and abruptly disappears at $K = 0.1883$ by the Potts renormalization-group flow mechanism generalized to quenched random interactions, namely by the unstable fixed distribution of the phase boundary between the two spin-glass phases and the stable fixed distribution sink of the weak-coupling spin-glass phase (Fig. 6) merging and annihilating.  Thus, for $K > 0.1883$, only the strong-coupling spin-glass phase occurs between the ferromagnetic and antiferromagnetic phases.}
\end{figure}
\begin{figure}[ht!]
\centering
\includegraphics[scale=0.25]{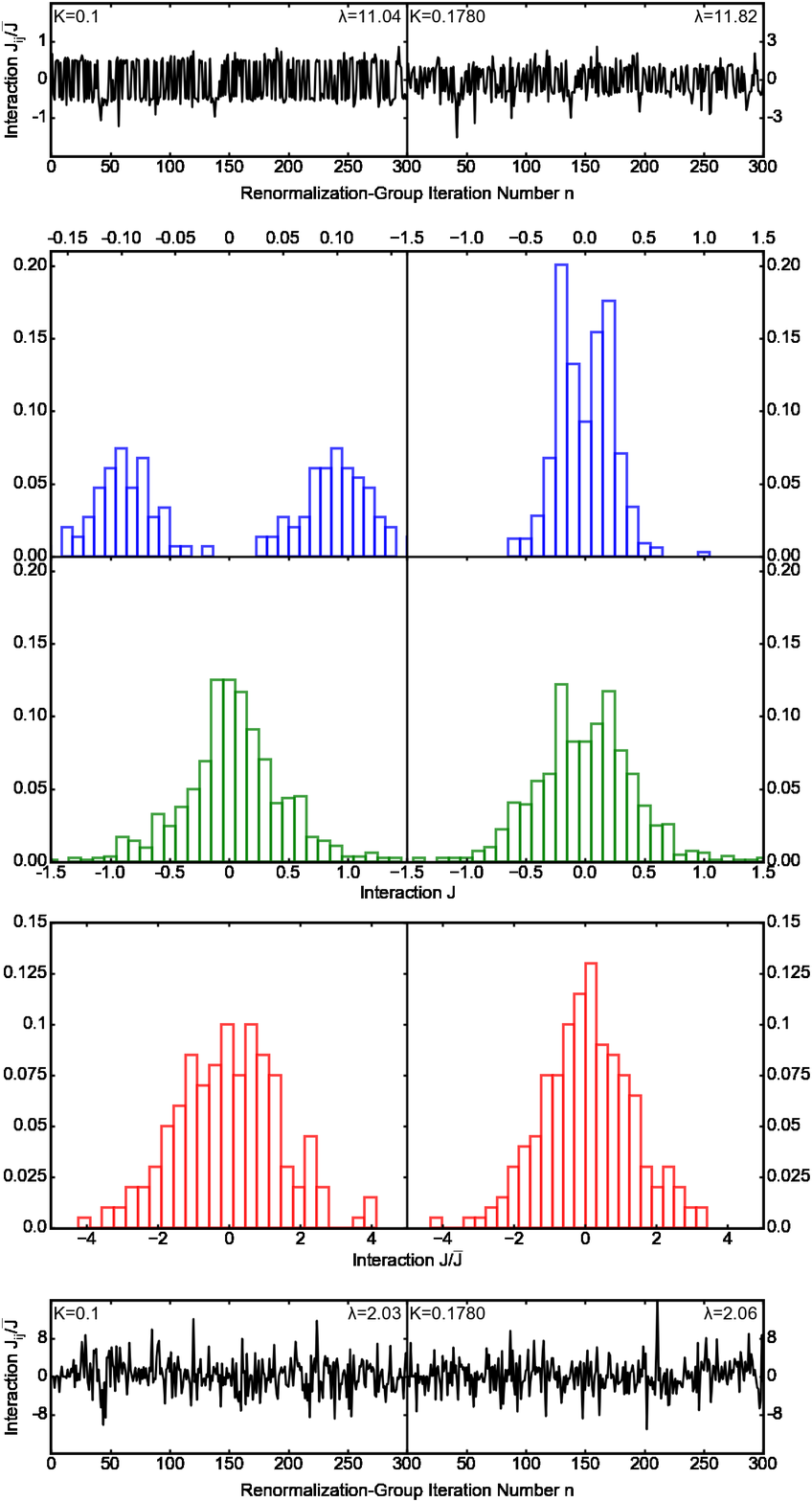}
\caption{Fixed distributions for the Ising spin glass with long-range spin-glass interaction $\pm K$ in $d=3$.  The left and right columns are for $K=0.1$ and 0.1780 respectively.  The top row gives the stable fixed distribution, i.e., sink, for the weak-coupling spin-glass phase.  The bottom  row gives the stable fixed distribution, i.e., sink, for the strong-coupling spin-glass phase.  The middle row gives the unstable fixed distribution for the phase transition between the weak- and strong-coupling spin-glass phases.  At the very top are the Lyapunov exponents for the weak-coupling sinks.  At the very bottom are the Lyapunov exponents for the strong-coupling sinks.}
\end{figure}
\section{Long-Range Spin-Glass Interactions and Spin-Glass-to-Spin-Glass Phase Transitions}
Calculated phase diagrams of the Ising spin glass with long-range spin-glass interaction $\pm K$ in $d=3$ are given in Fig. 5. From top to bottom, the phase diagrams are for $K=0.1,0.4,0.8$. The ferromagnetic and antiferromagnetic phases are marked respectively as F and A.  Between these phases, for $K=0.1$, there are the weak-coupling and strong-coupling spin-glass phases, respectively at high and low temperature.

Fixed distributions and chaos for these two spin-glasses with long-range spin-glass interaction $\pm K$ in $d=3$ are given in Fig. 6.  The left and right columns are for $K=0.1$ and 0.1883 respectively.  The top row gives the stable fixed distribution, i.e., sink, for the weak-coupling spin-glass phase.  The bottom row gives the stable fixed distribution, i.e., sink, for the strong-coupling spin-glass phase.  The middle row gives the unstable fixed distribution for the phase transition between the weak- and strong-coupling spin-glass phases.

As $K$ is increased, the stable sink fixed distribution for the weak-coupling spin-glass phase and the unstable fixed distribution for the phase transition approach each other, meaning perforce become identical (note the similarity the two distributions on the right top and bottom of Fig. 6, as compared with the left side), and annihilate each other, clearing the way for the renormalization-group flows to the strong-coupling spin-glass sink.  The weak-coupling spin-glass phase disappears and is replaced by the extended strong-coupling spin-glass phase, as seen for $K=0.4$ and 0.8 in Fig. 5.  This abrupt phase diagram change and its renormalization-group mechanism is the generalization to quenched random systems of the stable-unstable fixed-point annihilation (Fig. 3) of the Potts peninsular flow mechanism.

At the very top and bottom are the chaos and Lyapunov exponents for the weak-coupling and strong-coupling spin-glass phases.  Amazingly, as measured by the Lyapunov exponents, the weak-coupling spin-glass phase is more chaotic than the strong-coupling spin-glass phase.

We have also calculated phase diagrams of the Ising spin glass with decaying long-range spin-glass interaction $\pm K/r$, where $r$ is the separation between the spins in units of the nearest-neighbor separation in the original unrenormalized lattice.  As seen in Fig. 7, as $K$ is increased from 0, the strong-coupling spin-glass phase fully broadens becoming an intermediate phase between the ferromagnetic (antiferromagnetic) and disordered phases, and finally wholly replaces the disordered phase.
\begin{figure}[ht!]
\centering
\includegraphics[scale=0.17]{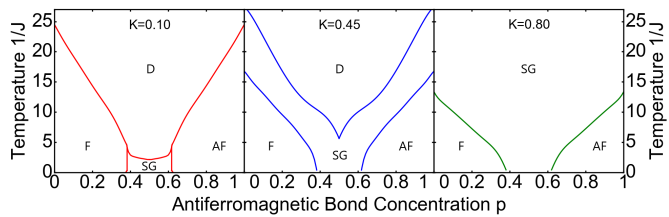}
\caption{Calculated phase diagrams of the Ising spin glass with decaying long-range spin-glass interaction $\pm K/r$, where $r$ is the separation between the spins in units of the nearest-neighbor separation in the original unrenormalized lattice.  The ferromagnetic (F), antiferromagnetic (A), strong-coupling spin-glass (SG), and disordered (D) phases are marked.  As $K$ is increased from 0, the strong-coupling spin-glass phase fully broadens becoming an intermediate phase between the ferromagnetic (antiferromagnetic) and disordered phases $(K=0.45)$, and finally wholly replaces the disordered phase $(K=0.80)$.}
\end{figure}
\section{Conclusion}
We have seen that the introduction, to the spin-glass system, of long-range ferromagnetic or spin-glass interactions reveal a plethora of new phases, spin-glass-to-spin-glass phase transitions, algebraic order, continuously varying runaway and non-runaway chaos, Potts-peninsular renormalization-group flows and precipitous phase diagrams, fixed-distribution annihilation.  The spin glasses are clearly a rich repository of complex-system behaviors.

\begin{acknowledgments}
Support by the Academy of Sciences of Turkey (T\"UBA) is gratefully acknowledged.
\end{acknowledgments}


\begin{references}

\bibitem{EdwardsAnderson} S.F. Edwards and P. W. Anderson, Theory of Spin Glasses, J. Phys. F {\bf 5}, 965 (1975).

\bibitem{Caglar1} T. \c{C}a\u{g}lar and A. N. Berker, Chiral Potts Spin Glass in d = 2 and 3 Dimensions, Phys. Rev. E {\bf 94}, 032121 (2016).
\bibitem{Caglar2} T. \c{C}a\u{g}lar and A. N. Berker, Devil’s Staircase Continuum in the Chiral Clock Spin Glass with Competing
Ferromagnetic-Antiferromagnetic and Left-Right Chiral Interactions, Phys. Rev. E {\bf 95}, 042125 (2017).
\bibitem{Caglar3} T. \c{C}a\u{g}lar and A. N. Berker, Phase Transitions Between Different Spin-Glass Phases and Between Different Chaoses
in Quenched Random Chiral Systems, Phys. Rev. E {\bf 96}, 032103 (2017).
\bibitem{Ostlund} S. Ostlund, Incommensurate and Commensurate Phases in Asymmetric Clock Models, Phys. Rev. {\bf24}, 398 (1981).
\bibitem{Surface3} M. Kardar and A. N. Berker, Commensurate-Incommensurate Phase Diagrams for Overlayers from a Helical Potts Model, Phys. Rev. Lett. {\bf 48}, 1552 (1982).

\bibitem{McKayChaos} S. R. McKay, A. N. Berker, and S. Kirkpatrick, Spin-Glass Behavior in Frustrated Ising Models with Chaotic Renormalization-Group Trajectories, Phys. Rev. Lett. {\bf 48}, 767 (1982).
\bibitem{McKayChaos2} S. R. McKay, A. N. Berker, and S. Kirkpatrick, Amorphously Packed, Frustrated Hierarchical Models: Chaotic Rescaling and Spin-Glass Behavior, J. Appl. Phys. {\bf 53}, 7974 (1982).
\bibitem{BerkerMcKay} A. N. Berker and S. R. McKay, Hierarchical Models and Chaotic Spin Glasses, J. Stat. Phys. {\bf 36}, 787 (1984).
\bibitem{Hartford} E. J. Hartford and S. R. McKay, Ising Spin-Glass Critical and Multicritical Fixed Distributions from a Renormalization-Group Calculation with Quenched Randomness, J. Appl. Phys. {\bf 70}, 6068 (1991).

\bibitem{ZZhu}Z. Zhu, A. J. Ochoa, S. Schnabel, F. Hamze, and H. G. Katzgraber, Best-Case Performance of Quantum Annealers on Native Spin-Glass Benchmarks: How Chaos Can Affect Success Probabilities, Phys. Rev. A {\bf 93}, 012317 (2016).
\bibitem{Katzgraber3}W. Wang, J. Machta, and H. G. Katzgraber, Bond Chaos in Spin Glasses Revealed through Thermal Boundary Conditions, Phys. Rev. B {\bf 93}, 224414 (2016).
\bibitem{Fernandez} L. A. Fernandez, E. Marinari, V. Martin-Mayor, G. Parisi, and D. Yllanes, Temperature Chaos is a Non-Local Effect, J. Stat. Mech. - Theory and Experiment, 123301 (2016).
\bibitem{Fernandez2} A. Billoire, L. A. Fernandez, A. Maiorano, E. Marinari, V. Martin-Mayor, J. Moreno-Gordo, G. Parisi, F. Ricci-Tersenghi, J.J. Ruiz-Lorenzo, Dynamic Variational Study of Chaos: Spin Glasses in Three Dimensions, J. Stat. Mech. - Theory and Experiment, 033302 (2018).
\bibitem{Wang2} W. Wang, M. Wallin, and J. Lidmar, Chaotic Temperature and Bond Dependence of Four-Dimensional Gaussian Spin Glasses with Partial Thermal Boundary Conditions, Phys. Rev. E {\bf98}, 062122 (2018).
\bibitem{Eldan} R. Eldan, A Simple Approach to Chaos For p-Spin Models, J. Stat. Phys. {\bf 181}, 1266 (2020).
\bibitem{Parisi3} M. Baity-Jesi, E. Calore, A. Cruz, L. A. Fernandez, J. M. Gil-Narvion, I. G.-A. Pemartin, A. Gordillo-Guerrero, D. I\~{n}iguez, A. Maiorano, E. Marinari, V. Martin-Mayor, J. Moreno-Gordo, A. Mu\~{n}oz-Sudupe, D. Navarro, I. Paga, G. Parisi, S. Perez-Gaviro, F. Ricci-Tersenghi, J. J. Ruiz-Lorenzo, S. F. Schifano, B. Seoane, A. Tarancon, R. Tripiccione, and D. Yllanes, Temperature Chaos Is Present in
    Off-Equilibrium Spin-Glass Dynamics, Comm. Phys. {\bf 4}, 74 (2021).

\bibitem{Aral} N. Aral and A. N. Berker, Chaotic Spin Correlations in Frustrated Ising Hierarchical Lattices, Phys. Rev. B {\bf 79}, 014434 (2009).

\bibitem{BerkerOstlund} A. N. Berker and S. Ostlund, Renormalisation-Group Calculations of Finite Systems: Order Parameter and Specific Heat for Epitaxial Ordering, J. Phys. C {\bf 12}, 4961 (1979).
\bibitem{Kaufman1} R. B. Griffiths and M. Kaufman, Spin Systems on Hierarchical Lattices: Introduction and Thermodynamic Limit, Phys. Rev. B {\bf 26}, 5022R (1982).
\bibitem{Kaufman2} M. Kaufman and R. B. Griffiths, Spin Systems on Hierarchical Lattices: 2. Some Examples of Soluble Models, Phys. Rev. B {\bf 30}, 244 (1984).

\bibitem{Hinczewski} M. Hinczewski and A. N. Berker, Inverted Berezinskii-Kosterlitz-Thouless Singularity and High-Temperature Algebraic Order in an Ising Model on a Scale-Free Hierarchical-Lattice Small-World Network, Phys. Rev. E {\bf 73}, 066126 (2006).
\bibitem{percolation} A. N. Berker, M. Hinczewski, and R. R. Netz, Critical Percolation Phase and Thermal BKT Transition in a Scale-Free Network with Short-Range and Long-Range Random Bonds, Phys. Rev. E {\bf 80}, 041118 (2009).
\bibitem{Jiang} K. Jiang, J. Qiao, and Y. Lan, Chaotic Renormalization Flow in the Potts model induced by long-range competition, Phys. Rev. E {\bf 103}, 062117 (2021).

\bibitem{Derevyagin2} G. Mograby, M. Derevyagin, G. V. Dunne,  and A. Teplyaev, Spectra of Perfect State Transfer Hamiltonians on Fractal-Like Graphs,J. Phys. A {\bf 54}, 125301 (2021).
\bibitem{Chio} I. Chio, R. K. W. Roeder, Chromatic Zeros on Hierarchical Lattices and Equidistribution on Parameter Space, Annales de l'Institut Henri Poincar\'{e} D, {\bf 8}, 491 (2021).
\bibitem{Teplyaev} B. Steinhurst and A. Teplyaev, Spectral Analysis on Barlow and Evans’ Projective Limit Fractals, J. Spectr. Theory {\bf 11}, 91 (2021).
\bibitem{Myshlyavtsev} A. V. Myshlyavtsev, M. D. Myshlyavtseva, and S. S. Akimenko, Classical Lattice Models with Single-Node Interactions on Hierarchical Lattices: The Two-Layer Ising Model, Physica A {\bf 558}, 124919 (2020).
\bibitem{Derevyagin} M. Derevyagin, G. V. Dunne, G. Mograby, and A. Teplyaev, Perfect Quantum State Transfer on Diamond Fractal Graphs, Quantum Information Processing, {\bf19}, 328 (2020).
\bibitem{Shrock} S.-C. Chang, R. K. W. Roeder, and R. Shrock, q-Plane Zeros of the Potts Partition Function on Diamond Hierarchical Graphs, J. Math. Phys. {\bf61}, 073301 (2020).
\bibitem{Monthus} C. Monthus, Real-Space Renormalization for Disordered Systems at the Level of Large Deviations, J. Stat. Mech. - Theory and Experiment, 013301 (2020).
\bibitem{Sariyer} O. S. Sar{\i}yer, Two-Dimensional Quantum-Spin-1/2 XXZ Magnet in Zero Magnetic Field: Global Thermodynamics from Renormalisation Group Theory, Philos. Mag. {\bf 99}, 1787 (2019).

\bibitem{AndelmanBerker} D. Andelman and A. N. Berker, Scale-Invariant Quenched Disorder and its Stability Criterion at Random Critical Points, Phys. Rev. B {\bf 29}, 2630 (1984).

\bibitem{Ilker1} E. Ilker and A. N. Berker, High q-State Clock Spin Glasses in Three Dimensions and the Lyapunov Exponents of Chaotic Phases and Chaotic Phase Boundaries, Phys. Rev. E {\bf 87}, 032124 (2013).
\bibitem{Ilker2} E. Ilker and A. N. Berker, Overfrustrated and Underfrustrated Spin Glasses in d=3 and 2: Evolution of Phase Diagrams and Chaos including Spin-Glass Order in d=2, Phys. Rev. E {\bf 89}, 042139 (2014).
\bibitem{Ilker3} E. Ilker and A. N. Berker, Odd q-State Clock Spin-Glass Models in Three Dimensions, Asymmetric Phase Diagrams, and Multiple Algebraically Ordered Phases, Phys. Rev. E {\bf 90}, 062112 (2014).

\bibitem{Thom} R. Thom, \textit{Stabilit\'{e} Structurelle et Morphongen\`{e}se} (Bunyamin, New York, 1972).

\bibitem{spinS7} B. Nienhuis, A.N. Berker, E.K. Riedel, and M. Schick, First- and Second-Order Phase Transitions in Potts Models: Renormalization-Group Solution, Phys. Rev. Lett. {\bf 43}, 737 (1979).
\bibitem{Nienhuis1} B. Nienhuis, E.K. Riedel, and M. Schick, Variational Renormalization-Group Approach to the q-State Potts Model in 2 Dimensions, J. Phys. A {\bf 13}, L31 (1980).
\bibitem{Nienhuis2} B. Nienhuis, E.K. Riedel, and M. Schick, Magnetic Exponents of the Two-Dimensional q-State Potts Model, J. Phys. A {\bf 13}, L189 (1980).
\bibitem{AndelmanPotts0} A. N. Berker, D. Andelman, and A. Aharony, 1st-Order and 2nd-Order Phase Transitions of Infinite-State Potts Models in One Dimension, J. Phys. A  {\bf 13}, L413 (1980).
\bibitem{AndelmanPotts1} D. Andelman and A. N. Berker, q-State Potts Models in d-Dimensions: Migdal-Kadanoff Approximation, J. Phys. A {\bf 14}, L91 (1981).
\bibitem{Nienhuis3} B. Nienhuis, E.K. Riedel, and M. Schick, q-State Potts Models in General Dimension, Phys. Rev. B {\bf 23}, 6055 (1981).

\bibitem{Kosterlitz} J. M. Kosterlitz and D. J. Thouless, Ordering, Metastability and Phase Transitions in 2 Dimensional Systems, J. Phys. C {\bf 6}, 1181 (1973).
\bibitem{Jose} J. V. Jos\'{e}, L. P. Kadanoff, S. Kirlpatrick, D. R. Nelson, Renormalization, Vortices, and Symmetry-Breaking Perturbations in 2-Dimensional Planar Model, Phys. Rev. B {\bf 16}, 1217 (1977).
\bibitem{BerkerNelson} A. N. Berker and D. R. Nelson, Superfluidity and Phase Separation in Helium Films, Phys. Rev. B {\bf 19}, 2488 (1979).
\bibitem{BerkerKadanoff1} A. N. Berker and L. P. Kadanoff, Ground-State Entropy and Algebraic Order at Low Temperatures, J. Phys. A {\bf 13}, L259 (1980).
\bibitem{BerkerKadanoff2} A. N. Berker and L. P. Kadanoff, Corrigendum, J. Phys. A {\bf 13}, 3786 (1980).

\bibitem{Collet} P. Collet and J.-P. Eckmann, \textit{Iterated Maps on the Interval as Dynamical Systems} (Birkh\"{a}user, Boston, 1980).
\bibitem{Hilborn} R. C. Hilborn, \textit{Chaos and Nonlinear Dynamics}, 2nd ed. (Oxford University Press, New York, 2003).

\bibitem{Demirtas} M. Demirtaş, A. Tuncer, and A. N. Berker, Lower-Critical Spin-Glass Dimension from 23 Sequenced Hierarchical Models, Phys. Rev. E 92, 022136 (2015).
\bibitem{ArtunBerker} E. C. Artun and A. N. Berker, Spin-s Spin-Glass Phases in the d=3 Ising Model, Phys. Rev. E {\bf 104}, 044131 (2021).

\end{references}
\end{document}